\documentclass{aastex63}

\received{???}
\revised{???}
\accepted{???}
\submitjournal{ApJ}

\usepackage{amsmath}

\shorttitle{Bastille Day Energetic Protons}
\shortauthors{Young et al.}
\watermark{Draft}
\graphicspath{{./}{figures/}}

\begin{document}

\title{Energetic Proton Propagation and Acceleration Simulated for the Bastille Day Event of July 14, 2000}

\correspondingauthor{Matthew A. Young}
\email{myoung.space.science@gmail.com}

\author[0000-0003-2124-7814]{Matthew A. Young}
\affiliation{University of New Hampshire \\
Morse Hall \\
8 College Road \\
Durham, NH, 03824, USA}

\author[0000-0002-3737-9283]{Nathan A. Schwadron}
\affiliation{University of New Hampshire \\
Morse Hall \\
8 College Road \\
Durham, NH, 03824, USA}

\author{Matthew Gorby}
\affiliation{University of New Hampshire \\
Morse Hall \\
8 College Road \\
Durham, NH, 03824, USA}

\author[0000-0003-1662-3328]{Jon Linker}
\affiliation{Predictive Science Inc. \\
9990 Mesa Rim Road \\
Suite 170 \\
San Diego, CA, 92121, USA}

\author[0000-0002-2633-4290]{Ronald M. Caplan}
\affiliation{Predictive Science Inc. \\
9990 Mesa Rim Road \\
Suite 170 \\
San Diego, CA, 92121, USA}

\author[0000-0003-1759-4354]{Cooper Downs}
\affiliation{Predictive Science Inc. \\
9990 Mesa Rim Road \\
Suite 170 \\
San Diego, CA, 92121, USA}

\author[0000-0003-3843-3242]{Tibor T{\"o}r{\"o}k}
\affiliation{Predictive Science Inc. \\
9990 Mesa Rim Road \\
Suite 170 \\
San Diego, CA, 92121, USA}

\author[0000-0002-1859-456X]{Pete Riley}
\affiliation{Predictive Science Inc. \\
9990 Mesa Rim Road \\
Suite 170 \\
San Diego, CA, 92121, USA}

\author[0000-0001-9231-045X]{Roberto Lionello}
\affiliation{Predictive Science Inc. \\
9990 Mesa Rim Road \\
Suite 170 \\
San Diego, CA, 92121, USA}

\author[0000-0001-7053-4081]{Viacheslav Titov}
\affiliation{Predictive Science Inc. \\
9990 Mesa Rim Road \\
Suite 170 \\
San Diego, CA, 92121, USA}

\author{Richard A. Mewaldt}
\affiliation{California Institute of Technology \\ 
Physics Department \\
Pasadena, CA, 91125, USA}

\author{Christina M. S. Cohen}
\affiliation{California Institute of Technology \\ 
Physics Department \\
Pasadena, CA, 91125, USA}



\begin{abstract}
This work presents results from simulations of the 14 July 2000 (``Bastille Day'') solar proton event. We used the Energetic Particle Radiation Environment Model (EPREM) and the CORona-HELiosphere (CORHEL) software suite within the SPE Threat Assessment Tool (STAT) framework to model proton acceleration to GeV energies due to the passage of a CME through the low solar corona, and compared the model results to GOES-08 observations. The coupled simulation models particle acceleration from 1 to 20 $R_\odot$, after which it models only particle transport. The simulation roughly reproduces the peak event fluxes, and timing and spatial location of the energetic particle event. While peak fluxes and overall variation within the first few hours of the simulation agree well with observations, the modeled CME moves beyond the inner simulation boundary after several hours. The model therefore accurately describes the acceleration processes in the low corona and resolves the sites of most rapid acceleration close to the Sun. Plots of integral flux envelopes from multiple simulated observers near Earth further improve the comparison to observations and increase potential for predicting solar particle events. Broken-power-law fits to fluence spectra agree with diffusive acceleration theory over the low energy range. Over the high energy range, they demonstrate the variability in acceleration rate and mirror the inter-event variability observed solar-cycle 23 GLEs. We discuss ways to improve STAT predictions, including using corrected GOES energy bins and computing fits to the seed spectrum. This paper demonstrates a predictive tool for simulating low-coronal SEP acceleration.
\end{abstract}

\keywords{}



\section{Introduction}
\label{sec:Introduction}
The acceleration of solar energetic particles (SEPs), and their transport to Earth and elsewhere in the heliosphere, has been recognized  since the 1940s \citep{Forbush_1946}.  Periods of enhanced energetic particle flux at 1 au present opportunities to study these fundamental processes, but also pose an increasing threat to geospace assets. Solar particle events (SPEs) can be hazardous to space-based crews in Earth orbit (e.g., aboard the International Space Station) and may imperil future crewed lunar or interplanetary missions \citep[][and references therein]{Cucinotta_etal-2010-SpaceRadiationRisk,Cucinotta_etal-2015-SafeDaysSpace}. They can even threaten aircraft communication and navigation systems, and increase long-term health risks for airline crews and passengers on polar flights. Therefore, providing a meaningful estimate of energetic particle flux at a particular location in the heliosphere is an important goal of current heliophysical research efforts. 

Large SPEs are typically associated with X-class solar flares and coronal mass ejections (CMEs) with speeds exceeding 1000 km s$^{-1}$, both of which typically arise from complex sunspot groups (also known as active regions) with strong magnetic fields. One mechanism for accelerating SEPs to the energies observed at Earth is via shocks --- or, more generally,  compressions --- that form low in the corona during the passage of a CME. After a compression forms, it propagates outward and accelerates particles over a finite space for a finite time, at which point magnetic connectivity and cross-field diffusion determine whether those particles arrive at an observer's location.

An analysis of Type-II radio bursts by \citet{Gopalswamy_etal-2005-TypeIIRadio} revealed that CME shocks responsible for many SEP events formed at a few solar radii above the solar surface, and that particle acceleration began at less than 10 solar radii ($R_\odot$). Later, \citet{Reames-2009-SolarEnergeticParticle,Reames-2009-SolarReleaseTimes} used velocity dispersion analysis (VDA; see \citet{Kahler_Ragot-2006-NearRelativisticElectron}) to derive solar particle release (SPR) times for 30 of the 45 GLEs during solar cycles 20-23. The results suggest particle release heights of approximately 2 $R_\odot$ for GLEs and above 3 $R_\odot$ for non-GLEs. They also suggest that shock acceleration occurs over a spatially broad region and occurs higher in the corona on the shock flanks. The notion of connecting an SPR time to an altitude above the solar surface extends back to \citet{Kahler-1994-InjectionProfilesSolar}, who also noted that SEP injection in three of the four analyzed GLEs resulted from a single CME-driven shock and likely did not occur at the center of the CME.

\citet{Zhang_etal-2001-OnTemporalRelationship} described a three-step process in which flare-associated CMEs are initiated in the low corona (1.3-1.5 $R_\odot$) and accelerated out to a few times $R_\odot$. Their work relied on four events with source regions close to the limb observed by the Large Angle and Spectrometric Coronagraph (LASCO) and EUV Imaging Telescope (EIT) aboard the Solar and Heliosphysics Observatory (SOHO). Detailed observations of so-called EIT waves, named after the instrument, prompted \citet{Liu_etal-2010-FirstSDOAIA} to propose that the associated sharp fronts likely reflect CME acceleration in the low corona, in support of the work by \citet{Zhang_etal-2001-OnTemporalRelationship}. Additional remote observations have solidified the picture of low-coronal shock formation \citep{Veronig_etal-2010-FirstObservationsDome} and their association with significant increases in proton flux at 1 au \citep{Kozarev_etal-2011-OffLimbSolar}.

\citet{Mewaldt_etal-2012-EnergySpectraComposition} showed that charge states observed for solar-cycle 23 GLE events require that the plasma density, $n$, and the particle acceleration time, $\tau$, obey the relation $n\tau \geq 7.5 \times 10^9$ s/cm${}^3$. That implies acceleration heights of around 1.5 $R_\odot$, which they found to be consistent with observations. They also note that acceleration along the CME flank, where particles have more access to higher density, would raise $n\tau$, making it easier to meet the minimum requirement.

Despite a wealth of energetic particle data at 1 au, our understanding of the origin of energetic particle events is limited by the historical paucity of observations closer to the Sun. While Parker Solar Probe and Solar Orbiter promise to provided valuable insight into the near-Sun environment, analysis of historical events must rely more heavily on remote observations, theory, and simulations. For example: A combination of 3-D modeling, remote observations, and 1-au data by \citet{Kouloumvakos_etal-2016-MultiViewpointObservations} indicated that previous SEP events might have been due to the first of two eruptive events in which the arrival of the shock at the STEREO-B footprint was consistent with the calculated SPR time, while the SPR time calculated for L1 was consistent with acceleration along the CME's western flank.

In this paper, we present results from simulations of the 14 July 2000 (``Bastille Day'') SPE using a software suite, called the SPE Threat Assessment Tool (STAT) framework \citep{Linker_etal-2019-CoupledMHDFocused}, that combines the Earth-Moon-Mars Radiation Environment Model \citep[EMMREM;][] {Schwadron_etal-2010-EarthMoonMars} and the CORona-HELiosphere \citep[CORHEL;][]{rileyetal2012}  software suite.
The tool enables us to model the proton acceleration process to GeV energies when a CME erupts, expands, and propagates through the low solar corona. This helps us understand how the acceleration process develops very close to the Sun --- a region where the interplay between CME initiation and expansion, the complex magnetic structures of the corona, and the expansion of the solar wind make energetic particle acceleration extremely difficult to accurately describe. Section \ref{sec:STAT framework} briefly describes the STAT framework and the components used in this investigations. Section \ref{sec:Event Background} describes the SPE event within the context of a multi-day period of solar activity; section \ref{sec:Results} presents the results of our study; section \ref{sec:Discussion} discusses the implications of our results and describes future avenues of research; and section \ref{sec:Conclusion} concludes the paper.

\section{STAT Framework}
\label{sec:STAT framework}

Modeling particle acceleration in the low coronal environment is especially challenging. The plasma and magnetic properties can vary by orders of magnitude, which leads to significant variation in the local Alfv\'en ($V_A$) and sound speeds ($C_S$). To investigate this complex but crucial region, STAT couples magnetohydrodynamic (MHD) simulations of CMEs in the low corona with 3D solutions of the focused transport equation for SEPs. It allows users to run the Energetic Particle Radiation Environment Module (EPREM, a component of EMMREM)  for precomputed Magnetohydrodynamic Algorithm outside a Sphere (MAS, a component of CORHEL) simulations of real CME events to simulate SEP events and provide diagnostics that can be compared with observations. STAT and its diagnostics have been recently described by \citet{Linker_etal-2019-CoupledMHDFocused}; here we briefly describe its primary  components.

\subsection{MAS Simulations}
\label{sec:MAS}

The properties of compressional regions such as shocks that drive SEP acceleration depend critically on the properties of the local plasma environment. To model a specific event, the simulation must realistically capture these properties for the time period under study. The MAS model has a long history of continued development and applications to this problem. While models with a simple energy equation can qualitatively reproduce coronal properties \citep{Mikic_Linker-1996-LargeScaleStructure,Linker_etal-1999-MagnetohydrodynamicModelingSolar,Mikic_etal-1999-MagnetohydrodynamicModelingGlobal} and are sufficient for exploring some dynamical aspects of boundary evolution \citep{Linker_etal-2011-EvolutionOpenMagnetic}, so-called thermodynamic MHD models \citep{Lionello_etal-2009-MultispectralEmissionSun,Riley_etal-2011-GlobalMHDModeling,rileyetal2012,Downs_etal-2013-ProbingSolarMagnetic,titovetal2017,linkeretal2017,mikicetal2018} are necessary to compute the plasma density and temperature with sufficient accuracy to simulate EUV and X-ray emission observed from space. In this approach, the energy equation accounts for anisotropic thermal conduction, radiative losses, and coronal heating. Inclusion of these extra physical terms is vital for obtaining realistic $V_A$ and $C_S$. 

To model a specific time period, a full-sun map of the photospheric magnetic field is obtained from an observatory or flux transport model and processed to create a boundary condition for the radial magnetic field  \citep[e.g.,][]{linkeretal2017}.  For this event, we developed a thermodynamic MHD simulation of the global corona using the procedure, equations, and coronal heating specification described by \citet{Lionello_etal-2009-MultispectralEmissionSun}; more recent MAS simulations use a Wave-Turbulence-Driven description of coronal heating  \citep[e.g.,][]{mikicetal2018}.  In the thermodynamic model, the temperature at the lower boundary is set to 20,000 K, similar to the upper chromosphere, and the upper boundary is at 20 solar radii ($R_\odot$), beyond the sonic and Alfv\'en critical points. 

CORHEL computes MAS solutions in the coronal and heliospheric domains separately.  Coronal solutions are used to provide the inner boundary condition for the heliospheric solutions for both steady-state background and dynamic CME simulations \citep{Lionello_etal-2013-MagnetohydrodynamicSimulationsInterplanetary}.  In principle, MHD solutions for both the coronal and heliospheric domains can be included in the EPREM calculation (described in section \ref{sec:EPREM}).  At the present time, STAT  employs only the MHD coronal domain within EPREM, and the remainder of the heliosphere is modeled with a simple spiral magnetic field created with a radially constant solar wind speed.  This restricts us to modeling the first few hours of an SEP event; once the CME leaves the coronal portion of the domain, possible SEP acceleration from the CME propagation in the heliosphere is not modeled.  STAT is presently being modified to incorporate both the coronal and heliospheric solutions in the EPREM simulations; these results will be the subject of future papers.

\subsection{EPREM Focused Transport Simulations}
\label{sec:EPREM}

EPREM models energetic particle acceleration and transport using a Lagrangian system, which co-moves with the plasma. To accomplish this, EPREM creates a spherical shell of simulation nodes at each time step and advances each node according to $\Delta \vec{r} = \vec{V}\Delta t$, where $\vec{r}$ is the 3-D node displacement, $\vec{V}$ is the 3-D flow velocity, and $\Delta t$ is the time step duration. It then calculates the distribution function for species $s$, $f_s(t, r, p, \mu)$, according to the focused transport equation,

\begin{eqnarray}
    \left[1 - \frac{\left(\vec{V}\cdot\hat{b}\right) v\mu}{c^2}\right]\frac{d f_s}{d t} & \qquad \textrm{(convection)} \nonumber \\
    + v\mu \hat{b} \cdot \nabla f_s & \qquad \textrm{(streaming)} \nonumber \\
    + \frac{\left(1 - \mu^2\right)}{2} \left[v \hat{b} \cdot \nabla \ln B - \frac{2}{v}\hat{b} \cdot \frac{d \vec{V}}{d t} + \mu\frac{d \ln \left(n^2 / B^3\right)}{d t}\right]\frac{\partial f_s}{\partial \mu} & \qquad \textrm{(adiabatic focusing)} \nonumber \\
    + \left[-\frac{\mu}{v}\hat{b} \cdot \frac{d \vec{V}}{d t} + \mu^2\frac{d \ln \left(n/B\right)}{d t} + \frac{\left(1 - \mu^2\right)}{2}\frac{d \ln B}{d t}\right]\frac{\partial f_s}{\partial \ln p} & \qquad \textrm{(cooling)} \nonumber \\
    = \frac{\partial}{\partial \mu}\left(\frac{D_{\mu\mu}}{2}\frac{\partial f_s}{\partial \mu}\right) + q\left(\vec{r}, p, t\right) & \qquad \textrm{(pitch-angle scattering and injection)}
    \label{eqn:focused_transport}
\end{eqnarray}

where $t$ is time, $r$ is distance along the stream line, $p$ is momentum, $\mu$ is the pitch-angle cosine, $\hat{b}$ is a unit vector parallel to the magnetic field $\vec{B}$, $v$ is the particle velocity (distinct from the flow velocity $\vec{V}$), $c$ is the speed of light, $q$ is the elementary charge, and $D$ is the diffusion tensor.

Each node advances outward with the solar wind flow and is linked to nodes on the neighboring shells. Each linked sequence of nodes defines a simulation stream representing a velocity path line --- the trajectory of fluid particles.  In steady-state (i.e., in the frame rotating with the Sun) these are also streamlines;  in places where the frozen-in assumption of ideal magnetohydrodynamics (MHD) holds, these lines also represent magnetic field lines.

The advantage of solving the transport problem in the co-moving frame is that it precludes the necessity of computing spatial gradients in flow velocity, which tend to introduce numerical errors that accumulate over many time steps. Instead, it requires the relatively simple task of computing the rates of change in plasma number density, $n$, and $\vec{B}$ at each stream node after being moved by a timestep. This methodology is based on the approach described in \citet{Kota_etal-2005-SimulationSEPAcceleration}, which follows from the theory developed by \citet{Skilling-1971-CosmicRaysGalaxy} and \citet{Ruffolo-1995-EffectAdiabaticDeceleration}. It was used by \citet{Kozarev_etal-2013-GlobalNumericalModeling} to study time dependent effects of solar energetic particle (SEP) acceleration in the low corona during CME evolution and by \citet{Schwadron_etal-2014-Synthesis3DCoronal} to model radiation doses at 1 au during a strong SPE event. In order to solve Equation \ref{eqn:focused_transport}, EPREM needs some model of $n$, $\vec{B}$, and $\vec{V}$ at each node. Simplified scenarios can use analytic forms of these plasma quantities but realistic modeling requires the use of MHD data such as that provided by CORHEL.

\section{Event Background}
\label{sec:Event Background}

The Bastille Day SPE occurred on 14 July 2000, during a four-day period of increased solar activity, including multiple CMEs, which produced high levels of energetic electrons and ions at 1 au \citep{Smith_etal-2001-ACEObservationsBastille}. This four-day activity period began on 12 July 2000 with a chromospheric H$\alpha$ flare observed by the \textit{Advanced Composition Explorer} (\textit{ACE}) spacecraft and ended on 16 July 2000 when energetic particle fluxes began to return to background levels. The solar eruption on 14 July 2000 was one of the largest of solar cycle 23 and produced one of that cycle's sixteen ground-level enhancement (GLE) events \citep{Bieber_etal-2002-EnergeticParticleObservations,Mewaldt_etal-2012-EnergySpectraComposition}.

\begin{figure}
    \centering
    \includegraphics[width=1.0\textwidth]{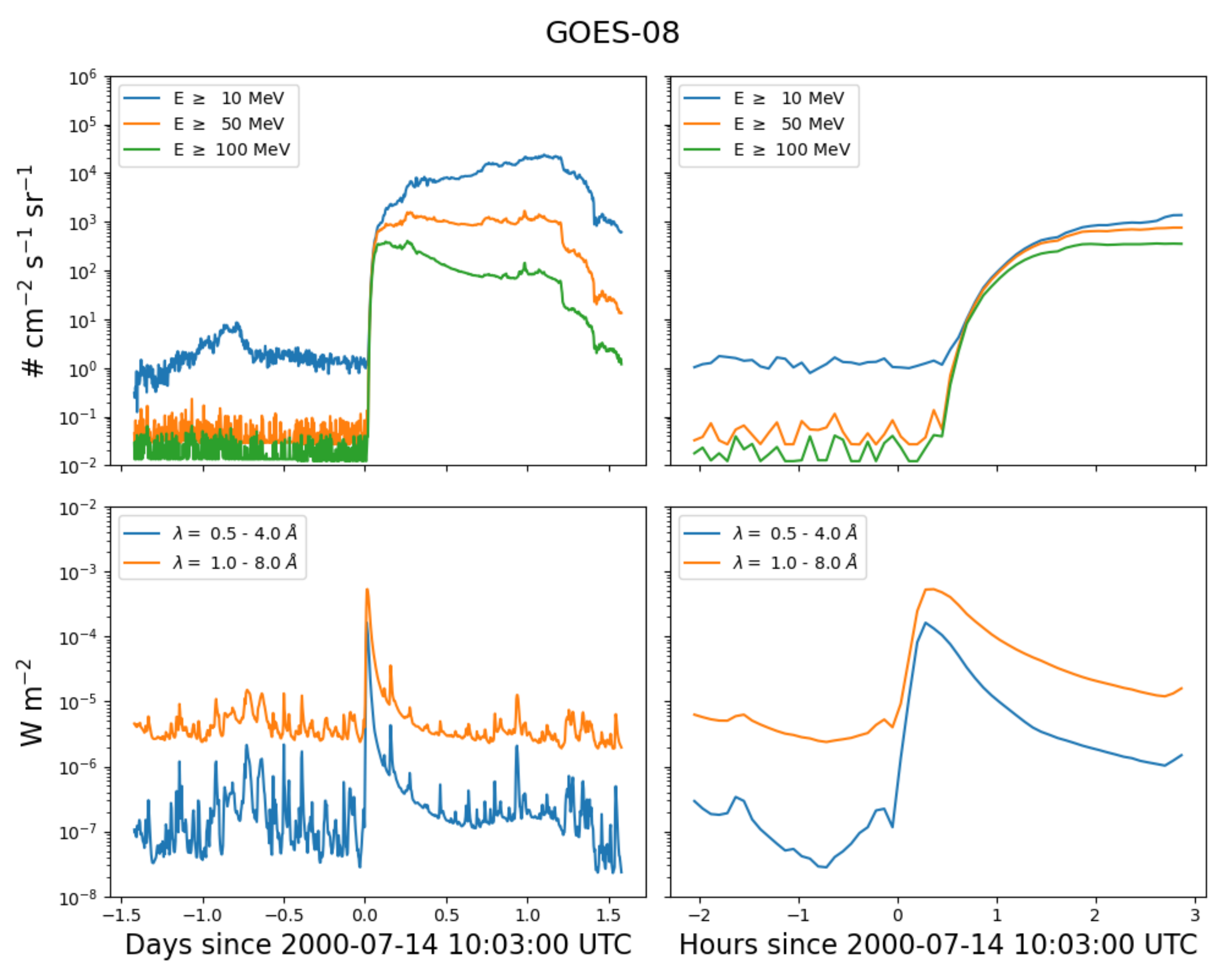}
    \caption{GOES-08 integral proton flux (top) and X-ray flux (bottom) during 13-16 July 2000 (left) and 08:00-13:00 UTC on 14 July 2000 (right). All times have been shifted to the event start at 10:03 UTC on 14 July 2000.}
    \label{fig:goes-protons_and_xrays}
\end{figure}

The solar eruption that precipitated the Bastille Day SPE originated in active region (AR) NOAA 9077. This active region produced an X-class flare starting at 10:03 UTC and peaking at 10:24 UTC on 14 July 2000. A halo CME associated with this flare subsequently became visible in the field-of-view of the Large Angle and Spectrometric Coronograph (LASCO) on board the Solar and Heliophysics Observatory (SOHO) at 10:54 UTC; initial speeds derived from second-order (i.e., constant acceleration) fits to four points on the CME's leading edge ranged from 1300-1700 km s$^{-1}$ \citep{Andrews-2001-LASCOEITObservations}. Figure \ref{fig:goes-protons_and_xrays} shows integral proton flux, in the top row, and X-ray flux, in the bottom row, observed by GOES-08 before and during the SPE. The left column shows data from 13-16 July 2000 and the right column shows data from 08:00-13:00 UTC on 14 July 2000. Time in both columns has been shifted so that the start of the event, at 10:03 on 14 July 2000, corresponds to 0.0 on all horizontal axes. The spike in X-ray flux marks the event start at 10:03 UTC, shortly after which proton integral flux rises sharply. 

\section{Results}
\label{sec:Results}

\subsection{MHD Results}

For our SPE modeling we utilize an MHD simulation of the 14 July 2000 Bastille Day eruption that was performed previously using CORHEL/MAS. Detailed descriptions of the simulation can be found in \cite{linker16} and \cite{Torok_etal-2018-SunEarthMHD}, here we restrict ourselves to the properties that are relevant to our investigation. 

\begin{figure}
\centering
\includegraphics[width=0.8\textwidth]{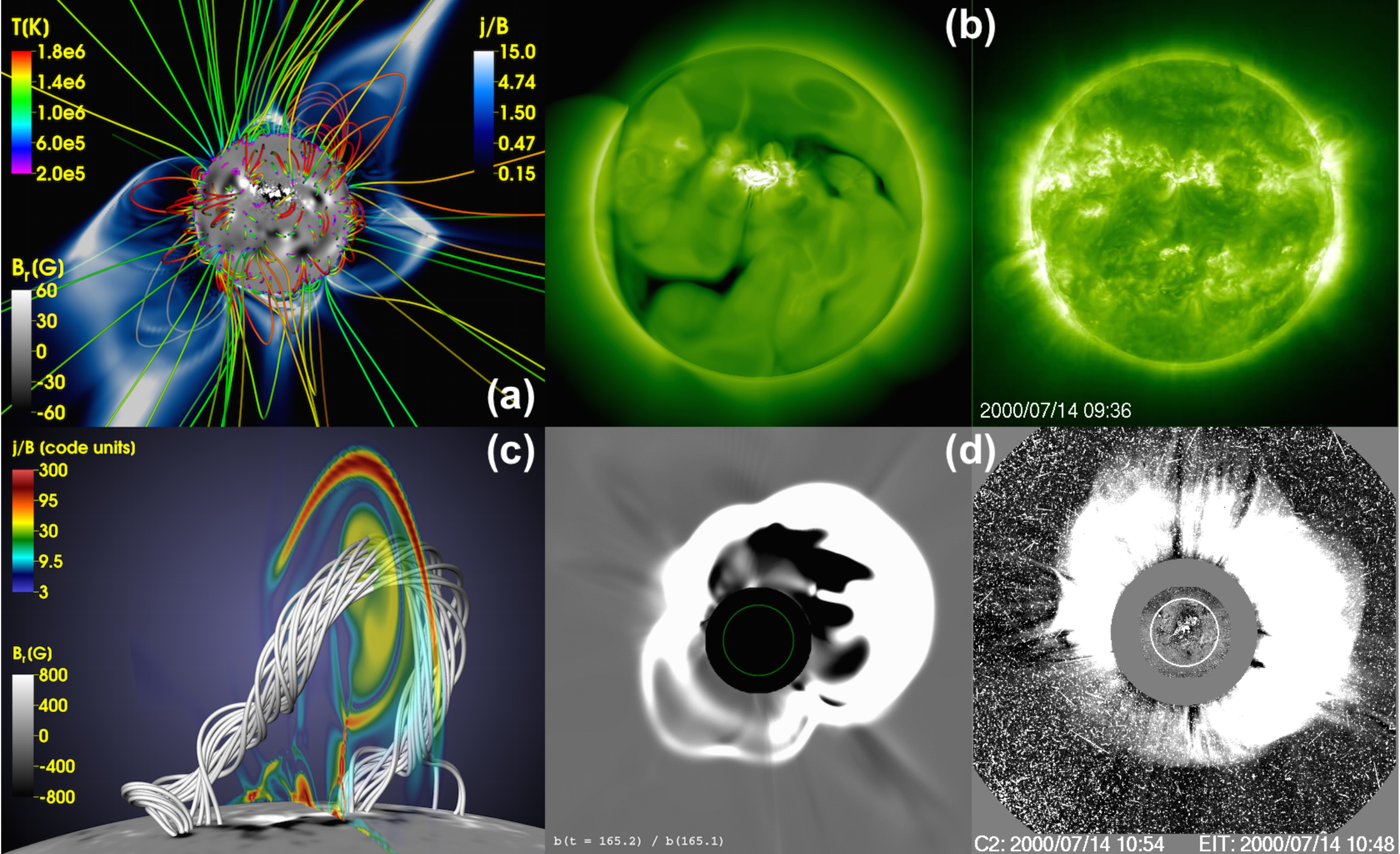}
\caption{
Snapshots of the MHD simulation.
(a) Coronal configuration seen from Earth at $\approx$\,09:30\,UT on 14 July 2000. 
NOAA AR 9077 is located north of disk center. Streamers are visualized by electric
currents; field lines are colored by temperature. The magnetogram, $B_r(R_\odot)$, 
is saturated at 60\,G. 
(b) Simulated and observed SOHO/EIT 195\,\AA\,emission of the corona shortly 
before the eruption. Emission from under-resolved ARs is not visible in the synthetic 
image. 
(c) Flux rope shortly after eruption onset. Electric currents show the compression 
regions and the shock in front of the rope.
(d) Simulated (running-ratio) and observed (difference) SOHO/LASCO C2 white-light 
image of the halo CME, about 40 minutes after eruption onset. The field-of-view is 
is 1.5-6\,$R_\odot$ in the synthetic image; the green circle marks the solar surface.
Adapted from \cite{Torok_etal-2018-SunEarthMHD}.
}
\label{fig:BD_simu}
\end{figure}

The simulation is conducted in several steps. First, a surface magnetogram 
is constructed and used to calculate a potential field that serves as the initial condition for the global magnetic field. This field is then subjected to a thermodynamic MHD relaxation 
(see Section \ref{sec:MAS})
until a steady-state plasma and magnetic field solution of the corona and solar wind is obtained (Figure\,\ref{fig:BD_simu}(a)), which can then be validated with observations (Figure\,\ref{fig:BD_simu}(b)). Next, in order to construct a current-carrying pre-eruptive configuration that can store the free magnetic energy required to power an eruption, a force-free magnetic flux rope is inserted along the polarity inversion line (PIL) of NOAA AR 9077. The highly elongated and curved rope is constructed using seven instances of the analytical TDm model \citep{titov14}, and relaxed towards a force-free state in a separate zero-$\beta$ MHD calculation (where thermal pressure and gravity are neglected). After the rope's insertion into the coronal solution and a further short relaxation, localized boundary flows converging towards the PIL are imposed, resulting in a slow rise of the flux rope, followed by its rapid acceleration and eruption, leading to a fast CME. 

Figure\,\ref{fig:BD_simu}(c) shows the core of the flux rope (white field lines) about two minutes after eruption onset. The shock front preceding the erupting rope (shown here via electric currents; cf. Figure\,\ref{fig:int_flux-goes-cme}) forms early on in the eruption, below $1.5\,R_\odot$. After its rapid acceleration low in the corona, the flux rope (or CME) reaches an almost constant propagation speed of $\approx 1500\,\mathrm{km\,s}^{-1}$, very similar to the observed speed (see Section\,\ref{sec:Event Background}). Figure\,\ref{fig:BD_simu}(d) shows a synthetic white-light image of the simulated halo CME, in good agreement with a SOHO/LASCO 2 observation at a corresponding time. 
We note that the coronal simulation was coupled to the heliospheric version of MAS and continued until the CME passed the Earth (see \citealt{Torok_etal-2018-SunEarthMHD} for details). \replaced{For the analysis described in this paper, only the coronal simulation was used.}{However, coupling to the heliospheric domain in EPREM is currently under development, so the analysis described in this paper used only the coronal portion of the MAS simulation.} 

\subsection{SEP Results}
\label{sec:SEP Results}

\begin{figure}
    \centering
    \includegraphics[width=1.0\textwidth]
    {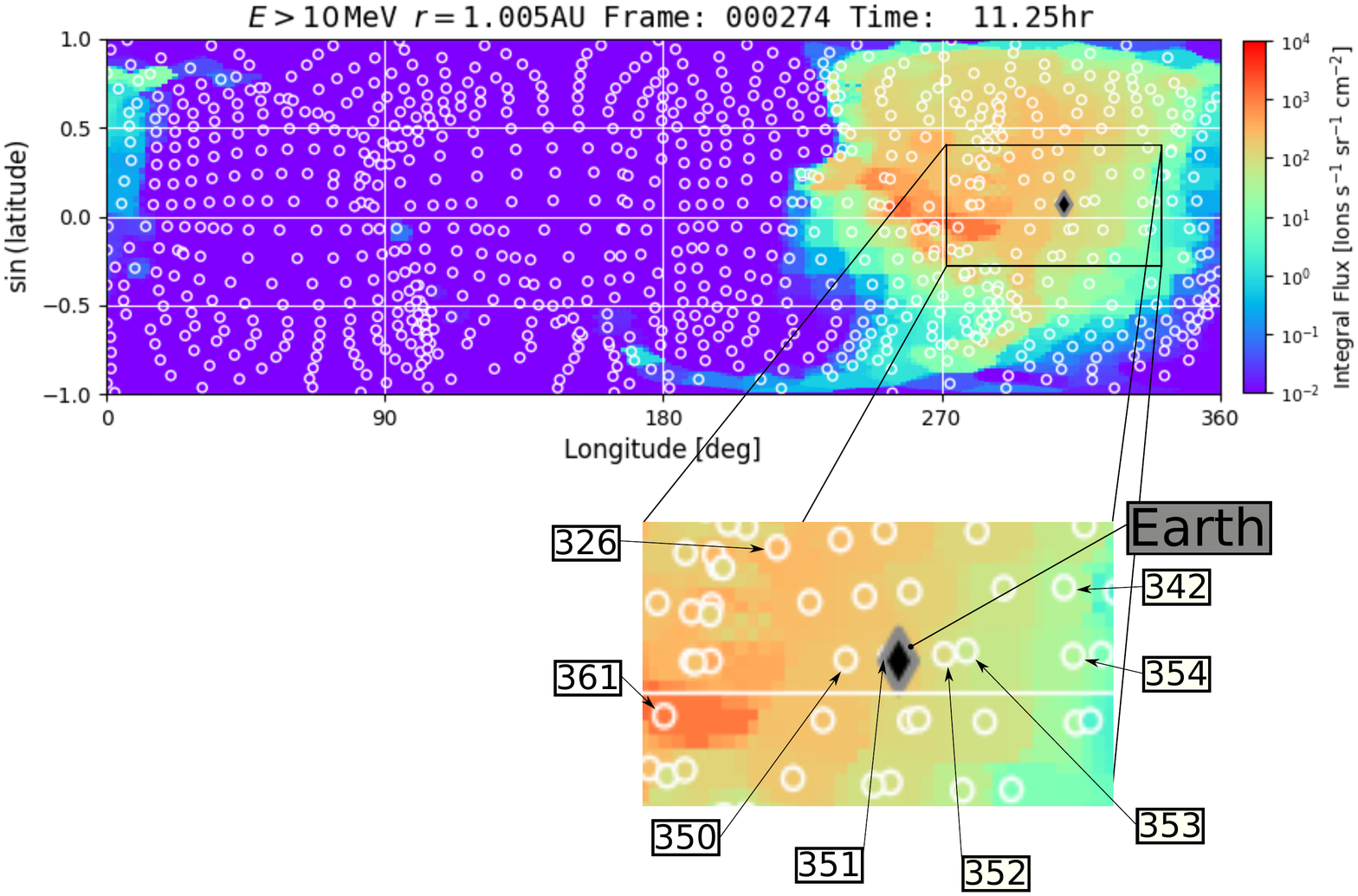}
    \caption{Integrated flux at 1 au from STAT output, with a call-out box showing streams (labeled by stream number) near Earth.}
    \label{fig:int_flux_hel_tp}
\end{figure}

Figure \ref{fig:int_flux_hel_tp} shows the distribution of EPREM nodes at 1 au, at the end of the simulation run. Color represents the amplitude of proton integral flux at energies above 10 MeV. Images like the top panel are part of the standard output of a STAT run. In both panels, a black diamond represents the approximate position of Earth at the current time step. The bottom panel zooms in on Earth's position and labels eight EPREM streams of interest: the four streams closest to Earth's position at approximately the same heliographic longitude (350-353), a stream near the center of the region of enhanced integral flux (326), a stream with extremely high integral flux (361), and two streams to the west of Earth (342 and 354) that are also near the edge of the SEP event. Stream 351 sits just behind the diamond and is therefore closest to Earth at the end of the simulation run, but stream 352 begins closest to Earth and remains closest during the development of the energetic particle event. From this point forward, figures will refer to this image and these stream numbers.

\begin{figure}
    \centering
    \includegraphics[width=1.0\textwidth]
    {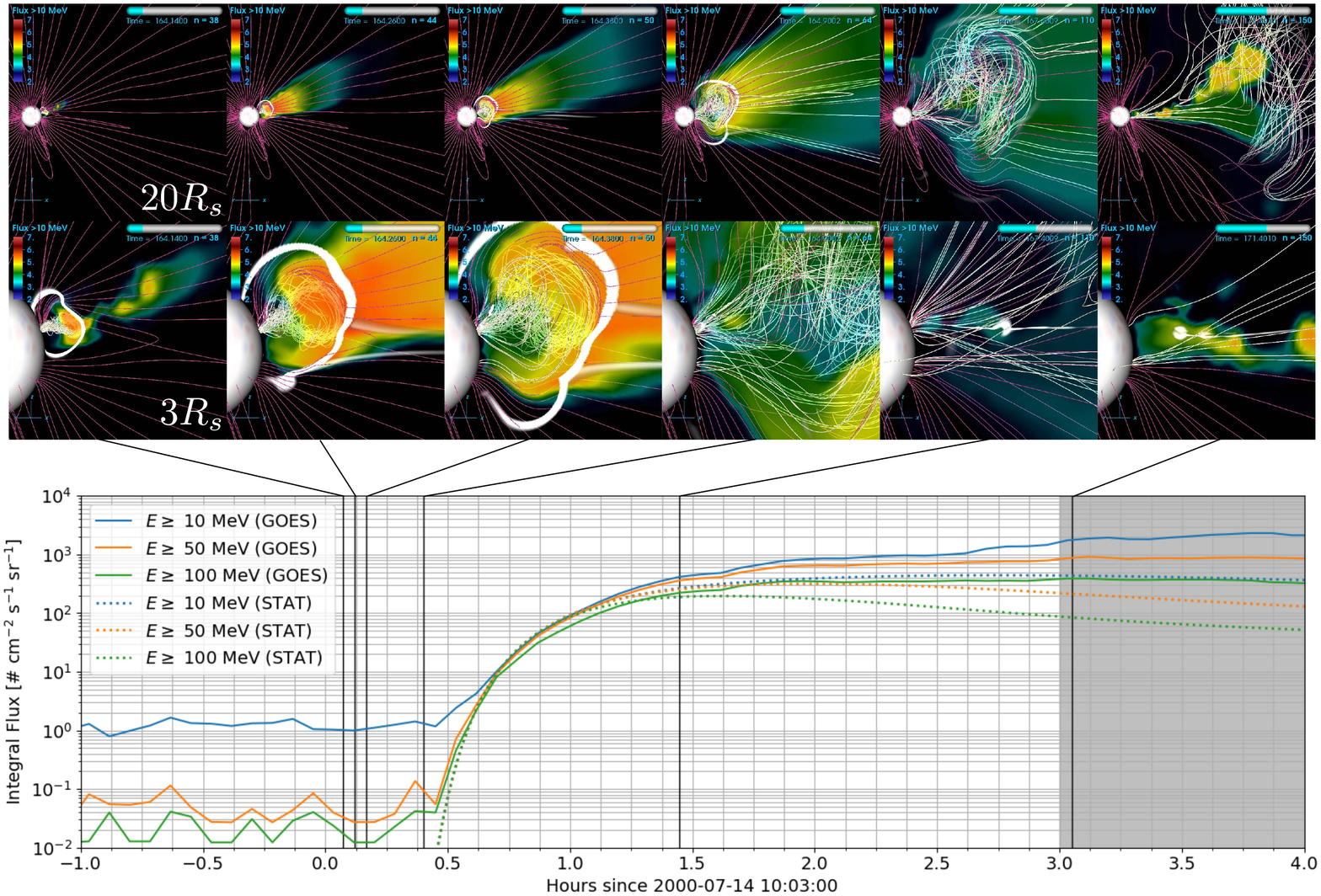}
    \caption{EPREM integral flux at 1 au on the stream nearest Earth during the early part of the simulation (352), compared to GOES-08. The sequences of panels in the upper rows show CME eruption and integral flux evolution at $20\ R_\odot$ and $3\ R_\odot$. Each panel corresponds to a time step as indicated by a vertical line in the plot in integral flux. \added{The shaded region beyond 3 hours after event start represents times after the CME left the coronal domain (i.e., the domain in which EPREM is coupled to MAS)}.}
    \label{fig:int_flux-goes-cme}
\end{figure}

STAT provides a global picture of integral proton flux in the standard reference bins of the NOAA GOES satellites at 1 au.  These are commonly used to assess energetic particle threats to space-based assets; STAT diagnostics allow inferences to be drawn about the physical drivers of these quantities.  
Figure \ref{fig:int_flux-goes-cme} shows integral proton flux in three energy channels --- $E \geq$ 10, 50, and 100 MeV --- on 14 July 2000. Solid lines represent GOES-08 proton data from the Energetic Particle Sensors (EPS) and dotted lines represent STAT output calculated from EPREM stream 352. The horizontal axis of the bottom panel represents time in hours since the start of the event at 10:03:00 UTC and the vertical axis represents integral proton flux in counts cm$^{-2}$ s$^{-1}$ sr$^{-1}$ \deleted{MeV$^{-1}$}. The two rows of color images above the line plot show meridional cuts at 3 and 20 $R_\odot$ from the CORHEL run, at the times indicated by vertical lines on the bottom plot. 

In each image, the gray sphere represents the solar surface, thin lines represent magnetic field lines, a thick white line indicates the peak in velocity compression ($-\nabla \cdot \vec{V}$), and the color scale represents the base-10 logarithm of integral flux at $E \geq 10$ MeV from 2 (blue) to 7 (red). The first three pairs of images capture the initial CME eruption out to approximately $3\ R_\odot$, before flux levels have risen above background at 1 au. Integrated flux begins to rise on this simulation stream around 20 minutes into the event, as the CME continues to expand and drive the compression outward, but the rise in GOES-08 data lags the simulation. We will address this discrepancy below. \replaced{The compression has considerably weakened by approximately 11:25:00 UTC (fifth panel) and the simulated integral flux has peaked around the level of the GOES-08 $E \geq 50$ MeV integral flux. After three hours, the compression region has passed $20 R_\odot$ and the simulated integral flux has begun to drop below GOES-08 levels.}{Simulated integral flux peaks around the level of GOES-08 between 1.0 and 1.5 hours after the event start, before flattening and beginning to fall off. This is due to the fact that the compression has considerably weakened by approximately 11:25:00 UTC (fifth panel) and has begun to leave the coronal domain. After three hours, the compression region has passed $20\ R_\odot$ and the simulated integral flux has noticeably dropped below GOES-08 levels.} It is important to note here that, though EPREM tracks protons out to 1 au, the MHD coupling in this simulation run extends only out to $20\ R_\odot$ (hereafter, the ``coronal domain''). Therefore, particle acceleration only occurs within the coronal domain. This paper will show that modeling acceleration within $20\ R_\odot$ captures many of the observed characteristics of this event, but the reader must bare in mind that results past roughly three hours after the start of the event do not account for additional acceleration between $20\ R_\odot$ and 1 au (hereafter, the ``heliospheric domain'').

\begin{figure}
    \centering
    \includegraphics[width=1.0\textwidth]{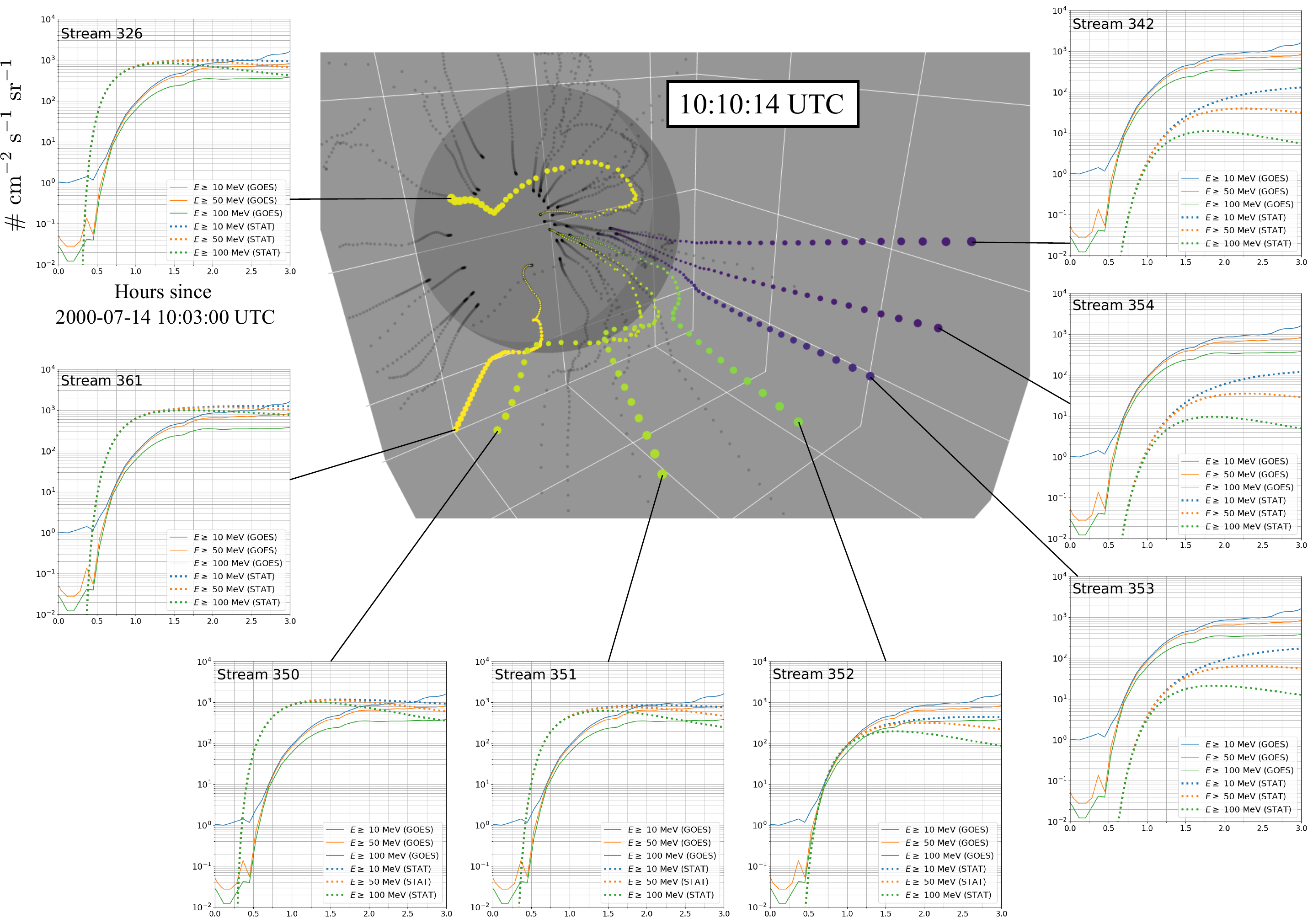}
    \caption{3-D projection of simulation streams during the CME eruption, with corresponding plots of integral flux at 1 au. Stream colors represent flux from low (dark) to high (light) along each stream, relative to the other streams pictured. The time shown is 7 minutes and 14 seconds after the start of the event. Streams 350-353 are each close to Earth at some point throughout the simulation. Stream 326 is near the center of enhanced integral flux. Stream 361 has extremely high integral flux. Streams 354 and 342 represent streams with relatively low integral flux. Nearby uncolored streams provide additional context. Line plots show integral flux at $E \geq 10$, 50, and 100 MeV observed by GOES-08 (solid lines) and on each stream (dotted lines). The horizontal and vertical axis labels shown for stream 326 apply to all streams. \added{An interactive version of the central panel is available at \url{https://prediccs.sr.unh.edu/sim/eprem/bastille-day-energetic-protons/}}}
    \label{fig:streams-3D-int_flux}
\end{figure}

Figure \ref{fig:streams-3D-int_flux} \explain{New version of figure, with bolder dashed lines} shows a subset of the EPREM simulation streams (i.e., linked EPREM nodes) at just over 7 minutes after the nominal event start. This UTC time corresponds to time step 44 in the simulation, during the early stage of the CME eruption. See the second pair of panels in the top rows of Figure \ref{fig:int_flux-goes-cme} for images of the CME eruption in the CORHEL simulation at the same time step. The larger, central panel of Figure \ref{fig:streams-3D-int_flux} displays eight streams in color: Streams 350, 351, 352, and 353, are each near the position of Earth (see Figure \ref{fig:int_flux_hel_tp}) at some point throughout the simulation, stream 326 sits near the center of the region of enhanced integral flux at 1 au, stream 361 has some of the highest integral flux values, and streams 354 and 342 have comparatively low integral flux values. The color of each stream represents the relative amount of flux it sees during the simulation, with dark colors representing relatively low flux and light colors representing relatively high flux. Since EPREM nodes initially trace out a Parker spiral, unperturbed streams are essentially radial at this distance. Meanwhile, streams perturbed by the CME drape around its path, thereby outlining its 3-D structure. The bends and kinks in these deformed stream lines turn out to be crucial in accelerating protons to SEP energies.

The smaller panels in Figure \ref{fig:streams-3D-int_flux} show simulated integral flux at 1 au on one of the highlighted streams against GOES-08 integral flux during the 14 July 2000 SEP event. The panel layouts are similar to the bottom plot in Figure \ref{fig:int_flux-goes-cme}. Simulated integral flux matches GOES-08 data to varying degrees at different points throughout the simulation. The onset and rise of integral flux along stream 352, which sits just to the right (heliographically west) of Earth at the end of the simulation, matches GOES-08 extremely well for about the first hour of the simulation run but turns over and falls below GOES-08 after the initial acceleration period. Stream 351, which ends up closest to Earth, shows integral flux increasing and rising before GOES-08 but decreasing to GOES-08 levels before falling below. Stream 350, which remains heliographically east of Earth for the duration of the simulation run, shows an even sharper rise and slightly higher peak amplitude compared to GOES-08. Near-Earth streams at higher and lower heliographic latitudes exhibit a longitudinal trend in integral flux similar to that of streams 350-353, as well as a slight north-to-south trend toward lower amplitudes. The overall northeast-to-southwest trend toward lower amplitudes mirrors the fact that Earth is in the lower right portion of the elevated integral flux in Figure \ref{fig:int_flux_hel_tp}; the additional east-to-west trend toward lower amplitudes and slightly slower rise times reflects both Earth's position relative to the integral flux enhancement as well as the streams' westward drift.

\begin{figure}
    \centering
    \includegraphics[width=1.0\textwidth]
    {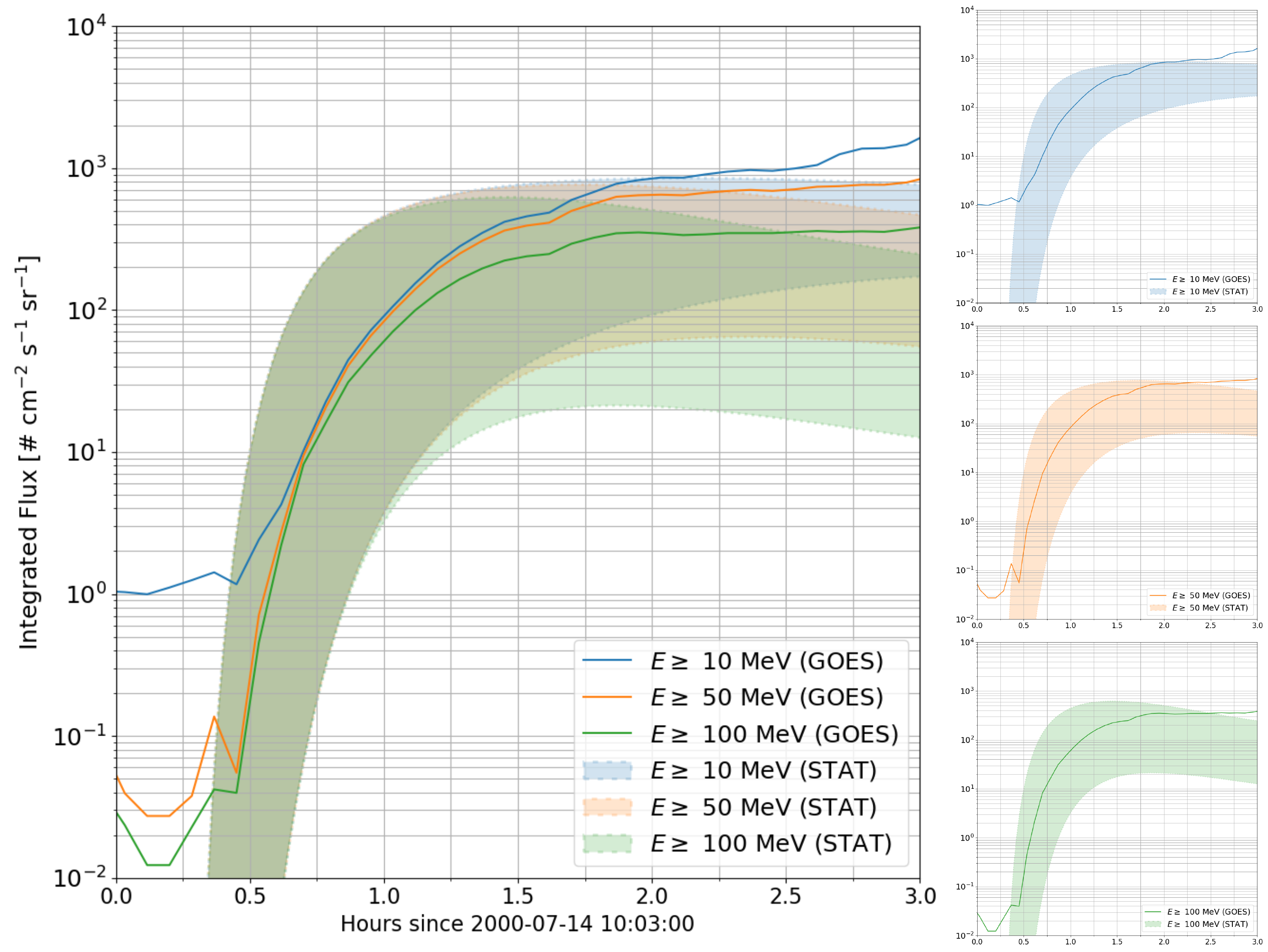}
    \caption{Envelope of EPREM integral flux at 1 au, on the three streams closest to Earth throughout the simulation run, compared to GOES-08. The integral flux envelope is defined as the area between the smallest and largest values (inclusive) among all three streams. The large panel shows GOES-08 and EPREM integral flux at $E \geq 10$, 50, and 100 MeV. Each panel in to the right shows one of the individual integral energy bins that make up the larger panel. As in Figure \ref{fig:streams-3D-int_flux}, the time axis spans the first three hours after 10:03 UTC.}
    \label{fig:int_flux-goes-envelope}
\end{figure}

The varying agreement with observations among streams is, in part, due to their relative motion over Earth's position. Figure \ref{fig:int_flux-goes-envelope} provides a rough accounting of the fact that streams move over Earth's position by showing the integral-flux envelope on streams 351, 352, and 353. The envelope is defined as the area between the minimum and maximum values (inclusive) among all three streams, at each time step. The three integral flux bins are color coded as in Figure \ref{fig:streams-3D-int_flux}, and the three-panel column on the right shows the individual contribution of each bin. Despite the variation in lower and upper bounds, as implied by Figure \ref{fig:streams-3D-int_flux}, the integral-flux envelope captures the rise and turn-over reasonably well, especially in the $E > 10$ MeV bin.

\begin{figure}
    \centering
    \includegraphics[width=1.0\textwidth]
    {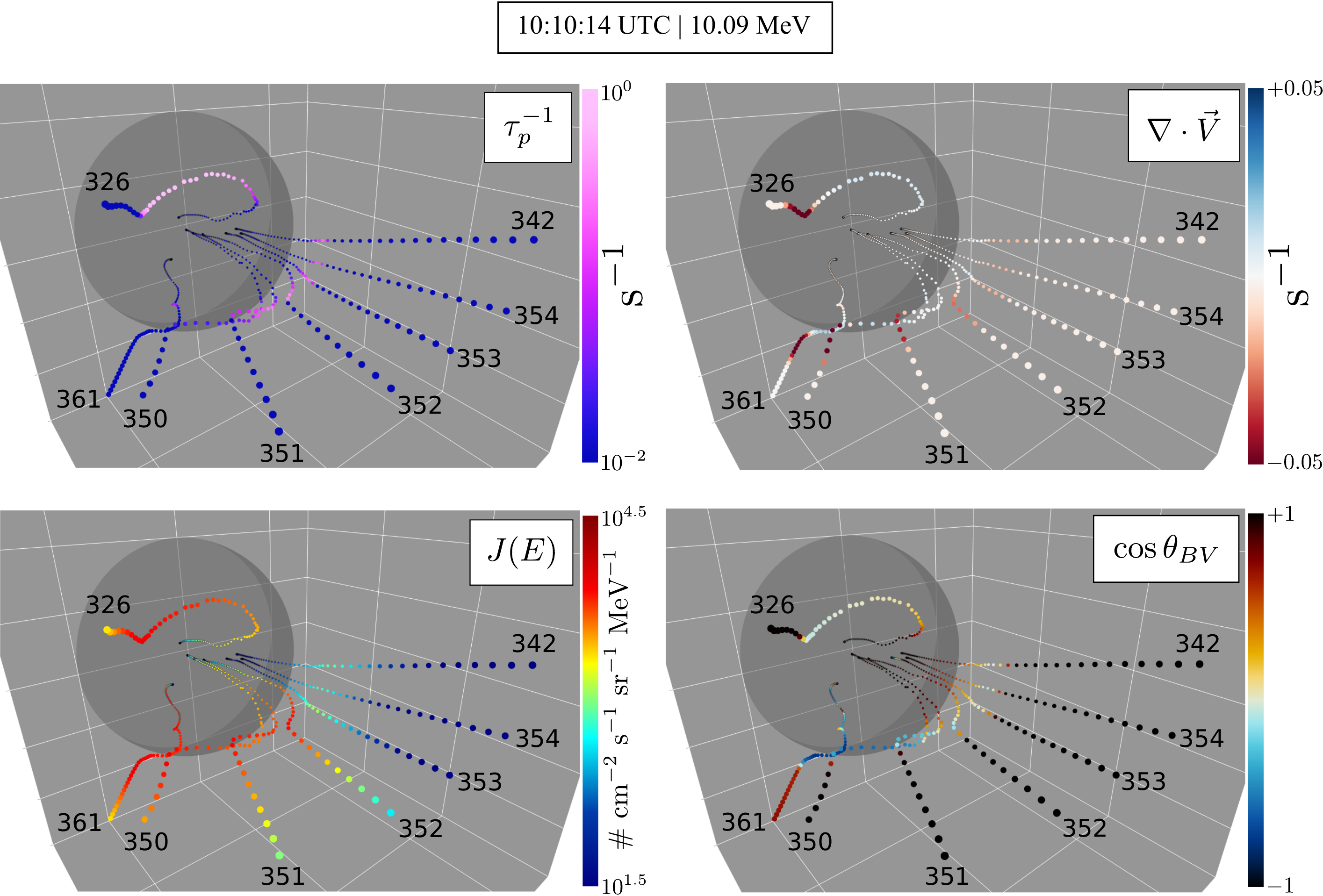}
    \caption{3-D view of nodes showing acceleration rate, $\tau_p^{-1}$, velocity divergence, $\nabla\cdot\vec{V}$, differential energy flux, $J(E)$, and cosine of the angle between the magnetic field and plasma flow, $\cos\theta_{BV}$, along the five streams indicated. The streams are shown at 10:10:14 UTC (7 minutes and 14 seconds after the start of the event). The values of $\tau_p^{-1}$ and $J(E)$ apply to protons with energy 10.09 MeV. An interactive version of each panel is available at \url{https://prediccs.sr.unh.edu/sim/eprem/bastille-day-energetic-protons/}}
    \label{fig:streams-3D-physics}
\end{figure}

Figure \ref{fig:streams-3D-physics} presents four versions of the same low-coronal view during the CME eruption as shown in Figure \ref{fig:streams-3D-int_flux}. Here, nodes along each stream are colored according to one of four physical quantities: The top left panel shows the local theoretical acceleration rate, $\tau_p^{-1}$ \citep{Schwadron_etal-2015-ParticleAccelerationLow}, of protons with $E \approx 10$ MeV, derived from MHD quantities along each stream. The top right panel shows velocity divergence $\nabla\cdot\vec{V}$ of the MHD flow. The bottom left panel shows the pitch-angle-averaged flux of protons with $E \approx 10$ MeV, computed from the simulated distribution function. The bottom right panel shows the cosine of the angle between the MHD magnetic and velocity fields.

Most streams with strong proton flux also have distinct bends where they drape around the flank of the CME (e.g., stream 326). Conversely, streams with little to no bend show weaker flux (e.g., stream 342). The regions of relatively large $-\nabla\cdot\vec{V}$, corresponding to the CME-driven shock, occur ahead of these stream-line bends whereas the peak in $\tau_p^{-1}$ occurs behind them. The shape of the CME suggested by the five stream lines suggests that there is considerable acceleration along the CME flank and that this acceleration is producing the flux observed at Earth. The change in $\cos\theta_{BV}$ from 1 to near or slightly below 0 indicates a change from field-aligned flow to cross-field flow and is approximately co-spatial with the region of peak $\tau_p^{-1}$. One exception to the correlation between stream-line bending and strong flux is stream 361, which traces more of a helical structure at this time step. Section \ref{sec:Discussion} will resolve this discrepancy and further discuss the implications of these results.

\begin{figure}
    \centering
    \includegraphics[height=0.8\textheight]
    {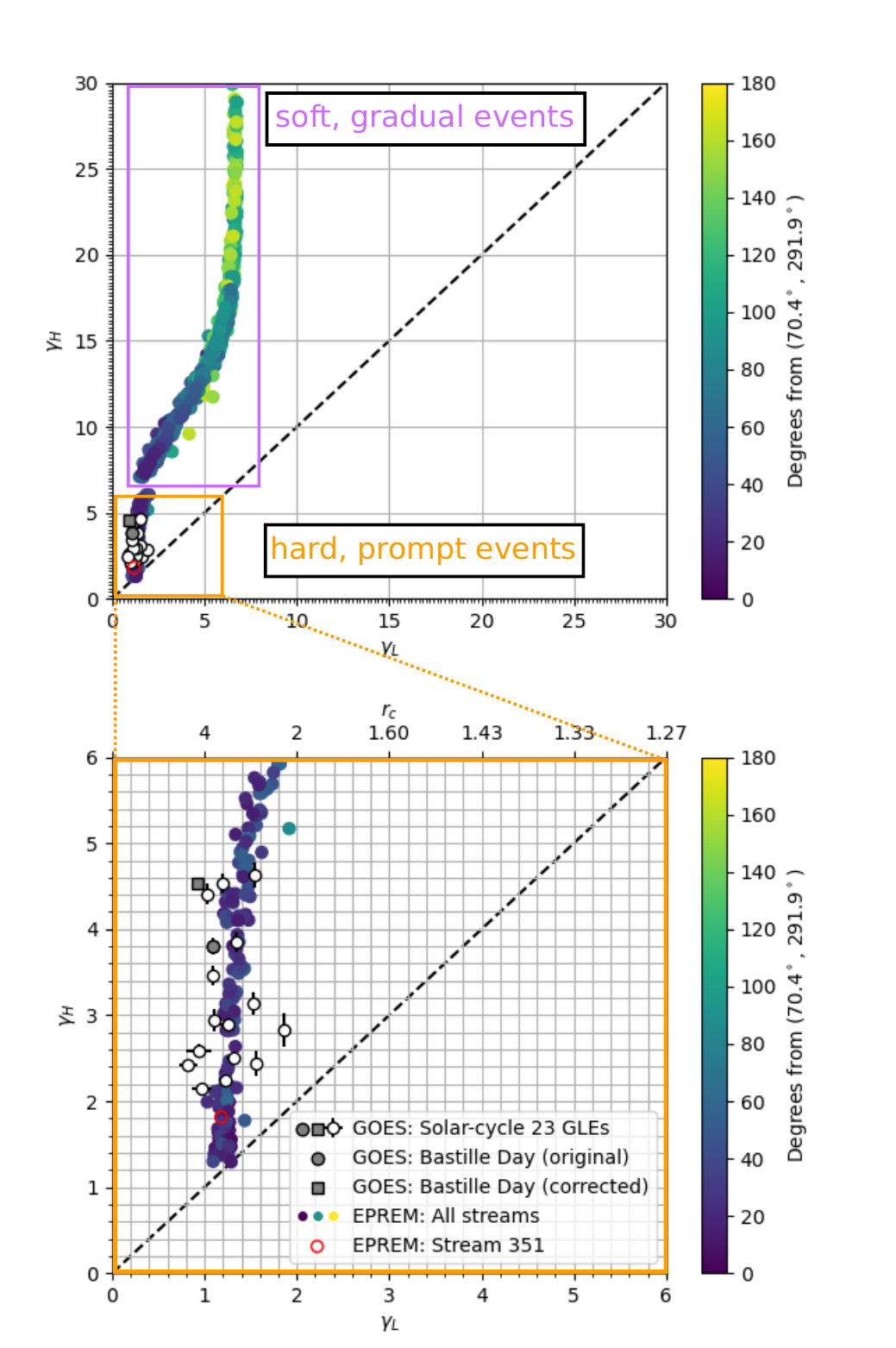}
    \caption{Low-energy power-law index, $\gamma_L$, versus high-energy power-law index, $\gamma_H$, for simulated and observed fluence spectra at 1 au. The bottom panel expands the lower left corner of the top panel. In both panels, filled circles represent spectral indices from all 16 GLEs observed by GOES during solar-cycle 23. The gray circle highlights the Bastille Day event. The gray square also corresponds to the Bastille Day GLE event, but was derived using corrected data (see Section \ref{sec:GOES energy bins}). The colored markers represent Bastille Day EPREM streams, with the color scale indicating distance from the center of elevated flux. The empty red circle corresponds to stream 351, which was near Earth. The top axis of the bottom panel lists the equivalent compression ratio, $r_c$, from diffusive shock acceleration theory for a given value of $\gamma_L$.}
    \label{fig:power-law_indices}
\end{figure}

Broken power laws are characterized by a power law with spectral index $\gamma_L$ over a low-energy domain, a power law with spectral index $\gamma_H > \gamma_L$ over a high-energy domain, and a break energy, $E_0$, separating the two domains. Proton fluence spectra observed at 1 au tend to exhibit broken power-law shapes due to the interplay between particle acceleration in the shock or compression ahead of a CME and the finite scales of the accelerator region. The former process controls the lower spectral index and the latter process controls the upper spectral index. \citet{Schwadron_etal-2015-ParticleAccelerationLow} showed theoretical fluence spectra with and without loss of particles due to a given flux bundle moving out of the acceleration region. They demonstrated that the spectrum above the break energy steepens (i.e., softens) when incorporating particle loss from the accelerator region since the effects of diffusion and convection can remove particles before they reach high energies. In their theoretical framework, $\gamma_L$ follows the prediction of diffusive shock acceleration (DSA), while $\gamma_H$ depends on magnetic rigidity, $R_g$. The reason for the former is that DSA efficiently accelerates particles to energies in the MeV range at quasi-parallel shocks until the particles gain enough energy to escape along the magnetic field. The reason for the latter is that parallel diffusion depends on rigidity as $\kappa_\parallel = \kappa_{\parallel0}\left(R_g/R_{g0}\right)^\chi$, where $R_{g0}$ is a reference rigidity.

Figure \ref{fig:power-law_indices} shows spectral indices in the high-energy subdomain versus spectral indices in the low-energy subdomain from power-law fits to simulated and observed fluence at 1 au. The color scale represents distance from the center of the region of enhanced integral flux, taken here to be stream 326, which is at $\left(\theta, \phi\right)$ = $\left(70.4^\circ, 291.9^\circ\right)$ (see Figure \ref{fig:int_flux_hel_tp}). Stream 351, which is approximately co-located with Earth at the final time step, is $18^\circ$ from stream 326. The upper horizontal axis on the bottom panel lists the compression ratio predicted by DSA for the corresponding lower power-law index, $r_c = 2(\gamma_L + 1)/(2\gamma_L - 1)$. A dashed line indicates $\gamma_L = \gamma_H$. 

The top panel of Figure \ref{fig:power-law_indices} shows values up to 30 for each spectral index. The dark-to-light trend in color as $\gamma_H$ increases implies that spectra quickly steepen in the high-energy subdomain as distance increases away from the region of enhanced integral flux. Note that this vertical trend continues above $\gamma_H = 30$. The spectral-index pairs shown in the top panel fall into two groups, labeled ``soft, gradual events'' and ``hard, prompt events''. Section \ref{sec:Discussion} will further discuss these groups. The bottom panel corresponds to the bottom-left region of the left panel, outlined in orange, extending to 6 in both axes. A substantial fraction of these points fall within $45^\circ$ of the reference point at stream 326.

Both panels highlight additional relevant points in $\gamma_H-\gamma_L$ space (see legend on bottom panel): Filled circles mark the spectral indices observed by GOES-08 during the 16 ground-level events (GLEs) of solar cycle 23 as reported by \citet{Mewaldt_etal-2012-EnergySpectraComposition}, with the Bastille Day (GLE) shown in gray and all others shown in white. A gray filled square marks spectral indices computed for this work using corrected GOES-08 fluence data -- see Section \ref{sec:GOES energy bins} for an explanation of this corrected data. Finally, an empty red circle indicates the EPREM stream nearest Earth.

\section{Discussion}
\label{sec:Discussion}

Here we discuss the implications of the results presented in Section \ref{sec:Results}. Specifically, the physical quantities presented in Figure \ref{fig:streams-3D-physics} and what they indicate about low-coronal acceleration, the distribution of spectral-index pairs in Figure \ref{fig:power-law_indices}, and potential sources of uncertainty in predictions based on our results.

\subsection{Acceleration rates}
\label{sec:Acceleration rates}

\citet{Kota_etal-2005-SimulationSEPAcceleration} noted that the expanding CME causes magnetic field lines to bend and compress, leading to two types of possible acceleration sites: density jumps at the parallel shock and magnetic field amplification just behind the shock. The images of stream 351 shown in Figure \ref{fig:streams-3D-physics} suggest the latter form of acceleration, in which magnetic-field amplification due to localized deflections along the CME flank, rather than the strong plasma compression in front of the CME, drives the high flux of energetic protons at 1 au. The highest values of $\tau_p^{-1}$ along this stream are $\sim 0.1$ s$^{-1}$.

The analysis presented in \citet{Schwadron_etal-2015-ParticleAccelerationLow}, which used the results of previous MAS simulations of a CME, anticipated the results from the coupled simulations shown here. They predicted that compressions, including (but not limited to) shocks, can readily accelerate particles to tens of MeV or more in a matter of minutes or less, especially as the geometry approaches perpendicular. Figure \ref{fig:streams-3D-physics} clearly shows large acceleration rates (small acceleration times) near the westward flank of the simulated CME, where the flow is far from parallel to the magnetic field. These extremely short acceleration times -- down to a few seconds on the stream with the highest proton flux -- rival those typically found in flare acceleration. The relationship between velocity and magnetic field in the MHD simulation, as captured by $\cos\theta_{BV}$, indicates that significant particle acceleration occurs where the field lines drape across the CME flank, rather than solely at the nose of the CME. The fact that $\tau_p^{-1}$ and $J(E)$ mirror the behavior of $\cos\theta_{BV}$ more closely than that of $\nabla\cdot\vec{V}$ further suggest that the relatively weak compressions on the CME flank are capable of accelerating protons to energies of tens to hundreds of MeV.

The case of stream 361 in Figure \ref{fig:streams-3D-physics} may appear, at first, to contradict the argument that stream-line (or field-line) draping along the flank of a CME produces strong particle acceleration in the low corona. In fact, it simply serves as a reminder that there are multiple ways to accelerate solar-wind particles to high energies. The lower apparent acceleration rate along stream 361 is partially due to the effect of projecting dynamic 3-D data onto a static 2-D image but it is true that its peak values of $\tau_p^{-1}$ at the time step shown are smaller than even those of stream 353. However, it nevertheless produces significant low-coronal flux and extremely high 1-au integral flux. The resolution of this discrepancy lies in the fact that the region of large $-\nabla\cdot\vec{V}$ coincides with a sharp transition from $\cos\theta_{BV} < 0$ to $\cos\theta_{BV} > 0$, indicating the location of a quasi-parallel shock. Analysis of the density compression ratio along this stream (not shown) suggests that a strong shock is directly responsible for accelerating protons along this stream via classical DSA

\added{We have examined similar figures for 100-MeV protons and found that the behavior differs in two notable ways: First, differential flux of 100-MeV protons decreases faster with distance from the CME than differential flux of 10-MeV protons. This implies that magnetic connectivity to the erupting CME plays a stronger role at higher energies, as observed by \citet{Gopalswamy_etal-2014-MajorSolarEruptions}. Second, 100-MeV protons along stream 326 have a peak flux value that is significantly higher than along the other streams shown in Figures \ref{fig:streams-3D-int_flux} and \ref{fig:streams-3D-physics}. Contrast this with Figure \ref{fig:streams-3D-int_flux}, in which 10-MeV protons along streams 326 and 361 have similar peak flux values. Recalling that stream 326 has a significant bend while stream 361 does not, this energy-dependent difference in where the peak flux emerges suggests that strong bends in field lines are responsible for producing the high-energy protons observed during GLEs. The interested reader will find interactive figures showing relative peak flux of 10-MeV and 100-MeV protons at \url{https://prediccs.sr.unh.edu/sim/eprem/bastille-day-energetic-protons/}}

\subsection{Broken power laws}
\label{sec:Broken power laws}

A second result from \citet{Schwadron_etal-2015-ParticleAccelerationLow} was that low-coronal particle acceleration by compressions leads to broken power-law fluence spectra at 1 au. Our present work shows the consistency between simulated fluence at 1 au and broken power laws formed by diffusive acceleration in shocks or compressions, followed by trapping and further acceleration for a fraction of the originally accelerated distribution. 

A key feature of the bottom panel of Figure \ref{fig:power-law_indices} is that values of $\gamma_L$ are constrained to within $1 \leq \gamma_L \leq 2$, or $4 \geq r_c \geq 2$. The upper bound in $r_c$ represents the theoretical upper limit for strong shocks; the lower bound represents a weaker yet still relatively strong shock with respect to fast-forward shocks observed in the heliosphere \citep{Kilpua_etal-2015-PropertiesDriversFast}. Values of $\gamma_H$, on the other hand, exhibit a much wider spread, implying that there is a high degree of variability in the spatial and temporal scales of the coronal accelerator regions along the streams that produce the SEPs observed at 1 au. Furthermore, the spread in fluence \replaced{spectra}{spectral indices} measured by GOES-08 during the 16 GLEs of solar cycle 23 is consistent with the spread in EPREM fluences for this single simulated event. This latter point suggests that inter-event variability among strong SEP events may be due, at least in part, to intra-event variability as a function of heliographic latitude and longitude.

\added{We note that this discussion has not included any information about either observed or simulated amplitude of proton fluence at 1 au. The simulation used a rough estimate for the proton seed-spectrum amplitude (see Section \ref{sec:Seed-particle spectrum}), which prevents us from drawing meaningful conclusions based on simulated fluence amplitude at 1 au. Our fitting procedure computed amplitudes of approximately $5 \times 10^6$ counts cm$^{-2}$ sr$^{-1}$ MeV$^{-1}$, whereas the values reported by \citet{Mewaldt_etal-2012-EnergySpectraComposition} range from $4.49 \pm 0.03 \times 10^7$ to $2.68 \pm 0.17 \times 10^9$ counts cm$^{-2}$ sr$^{-1}$ MeV$^{-1}$. The limited coupling domain (see the description of Figure \ref{fig:int_flux-goes-cme} in Section \ref{sec:SEP Results}) likely led to nonphysically low fluence amplitudes over the course of the simulation.}

\citet{Desai_etal-2016-ChargeToMass} found that low-energy spectral slopes in H-Fe spectra from 46 SEP events had values in the range 0.1-3 while high-energy spectral slopes had values in the range 0.5-9, and that the high-energy slope was typically larger than the low-energy slope. The values of $\gamma_L$ and $\gamma_H$ in Figure \ref{fig:power-law_indices} exhibit a similar relationship to their findings. They also showed that the spectral slopes were independent of species during a given SEP event. This suggests that the results shown in Figure \ref{fig:power-law_indices} should apply to other species.

As noted in Section \ref{sec:Results}, the spectral-index pairs in the top panel of Figure \ref{fig:power-law_indices} fall into two groups: soft, gradual spectra, with $\gamma_H \gtrsim 6.5$, and hard, prompt spectra, with $\gamma_H \lesssim 6.5$. The spectral-index pairs in the bottom panel correspond to hard, prompt spectra; they naturally include all the GLEs since only sufficiently energetic particles produce the atmospheric secondaries necessary for ground-level detection \citep{Lopate-2006-FiftyYearsGround}. This clustering into two groups is a result of the processes that create different spectral forms in fluence at 1 au. Many of the soft, gradual spectra occur far from the region of enhanced flux and essentially represent background fluences. They develop extremely soft high-energy spectra simply because there are few or no particles with energies above $E_0 \approx 30$ MeV. The asymptotic trend in soft, gradual spectra toward $\gamma_L \approx 7$ suggests particle acceleration by compressions with $r_c$ as low as $16/13 \approx 1.23$. The hard, prompt spectra likewise cluster because of the physical processes that produce measurable energetic particle events: The field-line draping shown in Figure \ref{fig:streams-3D-physics} accelerates a significant fraction of available particles to high energies in general, but the specific variation in $\tau_p$ among streams along the CME flank leads to the observed variability in $\gamma_H$ at 1 au. At the same time, diffusive acceleration due to shocks and compressions produces relatively consistent values of $\gamma_L$. \added{The energy-dependent behavior in peak flux values described above supports this conclusion that the particle acceleration due to field-line draping naturally produces both high-energy protons and variable high-energy fluence spectra.}

\subsection{Uncertainty in predictions}
\label{sec:Uncertainty in predictions}
\subsubsection{GOES energy bins}
\label{sec:GOES energy bins}

\begin{figure}
    \centering
    \includegraphics[width=1.0\textwidth]
    {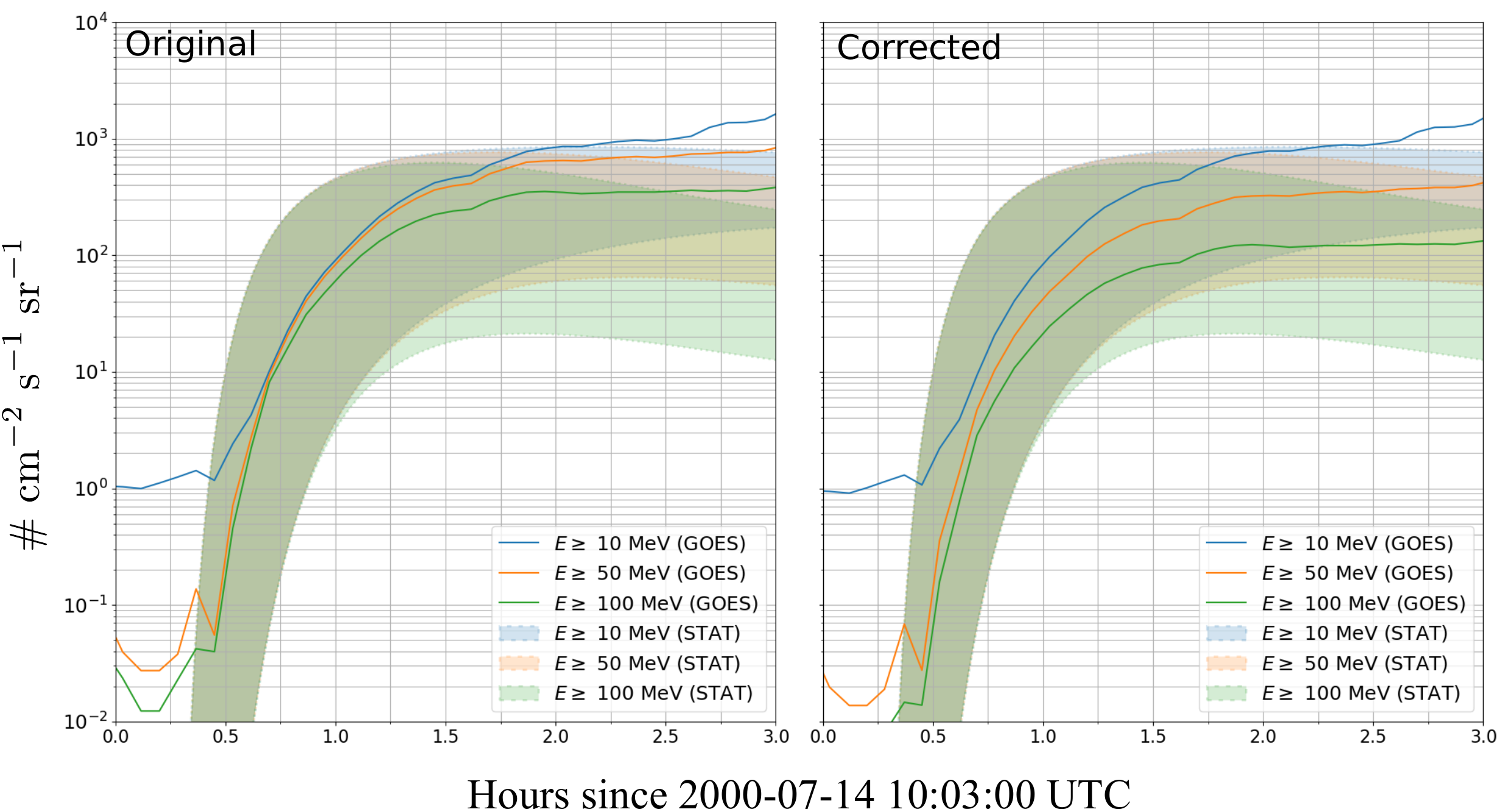}
    \caption{GOES-08 integral proton flux using original (left) and corrected (right) bins. The axis spans are identical to those in Figure \ref{fig:int_flux-goes-envelope}.}
    \label{fig:int_flux-orig-corr}
\end{figure}

The reliability of STAT predictions requires good agreement between simulation results and data. In most cases, one assumes that observations represent true values against which to judge simulation output. However, properly converting raw data into higher-level products requires care and an appropriate interpretation of the raw data. \citet{Sandberg_etal-2014-CrossCalibrationNOAA} used a novel intercalibration scheme to derive effective energy values for GOES/EPS bins, using the IMP-8/Goddard Medium Energy Experiment (GME) data set as a reference. See \citet{Cohen_Mewaldt-2018-GroundLevelEnhancement} for additional motivation behind deriving effective energy values and ranges. So far, this work has compared simulation results to the publicly available (uncorrected) GOES-08 data. Figure \ref{fig:int_flux-orig-corr} reproduces the integral flux envelope from Figure \ref{fig:int_flux-goes-envelope} (``Original'') alongside a similar plot that uses re-calibrated GOES-08 energy bins (``Corrected''). Due to the increased importance of corrections in higher energy bins, the effect is more pronounced in the $\geq 50$ and $\geq 100$ MeV channels. This effectively spreads out both the relative rise times and amplitudes of the three traces. When compared to corrected integral flux, the simulated integral flux envelope captures more of each GOES-08 trace, especially in the highest energy bin.

\subsubsection{Seed-particle spectrum}
\label{sec:Seed-particle spectrum}

The observed fluence depends on the form of the seed population through which the CME propagates. EPREM assumes a seed-particle spectrum that depends on energy, $E$, and radial distance from the Sun, $r$, as 
\begin{equation}
    f_{seed}(E,r) = \frac{m_p^2}{2E}\left(\frac{J_0}{\xi}\right)\left(\frac{r_1}{r}\right)^\beta\left(\frac{E}{E_r}\right)^{-\gamma}\exp\left(-\frac{E}{E_0}\right),
    \label{eqn:seed_spectrum}
\end{equation}
where $m$ is the particle mass, $J_0$ is the amplitude of the $^{4}$He fluence spectrum, $\xi$ is the ratio of $^{4}$He to H, $r_1$ is a fixed reference distance, $\beta$ is the radial scaling dependence, $E_r$ is a reference energy, $\gamma$ is the energy power-law dependence, and $E_0$ is the roll-over energy. Table \ref{tab:seed_spectrum} lists the parameter values used for this simulation run. \replaced{These values are estimates based on historical 1-au data and physical arguments.}{The values of $\xi$, $\gamma$, and $E_0$ were chosen based on the work described in \citet{Dayeh_etal-2009-CompositionSpectralProperties} \citep[See also][]{Kozarev_etal-2013-GlobalNumericalModeling}. The value of $J_0$ was a rough estimate based on averaged quiet-time GOES-08 proton flux before the event; the value of $E_r$ simply sets the reference point for this estimate.} Future simulation runs will use physical estimates of seed-population parameters derived from fits to 1-au satellite data (e.g., from GOES-08, ACE, Wind, and STEREO), then compare the results to simulated fluence and integral flux to improve predictions. Observations near the Sun by Parker Solar Probe and Solar Orbiter will also provide direct estimates of the SEP seed spectrum. \added{Improved estimates of the fluence amplitude, in particular, for a range events will greatly improve STAT as a predictive tool.}

\begin{deluxetable*}{cccl}
    \tablenum{1}
    \tablecaption{
    Seed spectrum parameters used in the simulation
    \label{tab:seed_spectrum}
    }
    \tablewidth{0pt}
    \tablehead{
    \colhead{Symbol} & \colhead{Value} & \colhead{Units} & \colhead{Description}
    }
    \startdata
    $J_0$ & 20.0 & cm$^{-2}$ s$^{-1}$ sr$^{-1}$ MeV$^{-1}$ & Fluence amplitude \\
    $\xi$ & 0.10 & -- & $^{4}$He/H \\
    $r_1$ & 1 & au & Reference distance from Sun \\
    $\beta$ & 1.7 & -- & Radial scaling dependence \\
    $E_r$ & 1.0 & MeV & Power-law reference energy \\
    $\gamma$ & 2.0 & -- & Power-law index \\
    $E_0$ & 1.0 & MeV & Roll-over energy
    \enddata
\end{deluxetable*}

\section{Conclusion}
\label{sec:Conclusion}

This paper has presented results from the Bastille Day energetic particle event on 14 July 2000. We used two core elements of the STAT simulation framework --- EPREM and CORHEL --- to model particle acceleration due to a CME-driven shock, and subsequent transport to 1 au. The Bastille Day event was one of the GLE events of solar cycle 23 and is an excellent example of low-coronal proton acceleration. Comparison between simulated and observed integral flux shows that, though acceleration in the coupled simulation occurs only within 20-30 $R_\odot$, simulated and observed integral flux at a single near-Earth observer show good agreement early in the event. Including multiple simulated near-Earth observers in a min-max envelope further improves the predictive capabilities of this tool. Simulated particle arrival times match observations very well despite the fact that the CME launch time was chosen based on EUV observations, rather than a fit to particle arrival times.

Broken power-law fits to simulated fluence spectra at 1 au yield a trend in low-energy spectral index versus high-energy spectral index that is consistent with observations: The low-energy spectral index is confined to a narrow range while the high-energy spectral index varies over a wide range. Values of the low-energy spectral index match predictions of diffusive acceleration theory for density compression ratios from 2-4, which are also observed in the low-coronal portion of simulated field lines. Values of the high-energy spectral index are determined by the spatial and temporal extent of low-coronal particle-acceleration regions; they exhibit much greater variation due to the low-coronal variation in density, velocity, and magnetic field during CME passage. The variability in high-energy spectral index of simulated fluence is consistent with variability in observed fluence during all 16 GLEs of solar cycle 23, suggesting that at least some historical \emph{inter}-event variability may be due to point observations of \emph{intra}-event variability.

The STAT framework is in active development, and we discussed two ways in which we may improve future predictions: comparing simulated integral flux to GOES data using corrected energy bins, and fitting the energetic-proton seed spectrum to fluence data. Motivation for the first improvement comes from published works that have demonstrated how novel intercalibration schemes for GOES energy bins can increase agreement with the IMP-8 reference data set. We found that 1-au integral flux from our simulation agreed better with the corrected GOES-08 data during the early part of the event, and no worse after the period of initial acceleration, than with the original GOES-08 data. The second proposed improvement seeks to refine our method for estimating EPREM input parameters related to the energetic-proton seed spectrum, which is currently based on historical estimates of the ${}^4$He spectrum and does not vary from event to event. We have begun by fitting the seed-spectrum functional form to averaged GOES-08 proton fluence data from a few hours before each event and will continue by incorporating additional data sets (e.g., ACE, Wind, and STEREO). Results will be the subject of future publications. An additional improvement to STAT's predictive capabilities that is in progress is a framework for coupling heliospheric MHD data from CORHEL into EPREM, to allow acceleration beyond the current 20-30 $R_\odot$ domain. This is critical for modeling a full event, which is a requirement in calculating radiation dose levels.

We have thus provided results of the SPE Threat Assessment Tool (STAT) framework for the onset of the Bastille Day Event of 2000. The simulation itself roughly reproduces the timing of the event, the peak event fluxes, and the spatial location of peak flux. The coupled EPREM-CORHEL model accurately describes the acceleration processes active in the low corona and resolves the sites of the most rapid acceleration close to the Sun. Disagreement between simulated and observed particle fluxes after about three hours post-onset are to be expected, since the coupling only extends to the outer edge of the coronal domain. This tool nonetheless demonstrates predictive capabilities that may be a key component for understanding the conditions associated with space radiation events and the hazards they pose to future space-based missions.

\acknowledgments
This work was supported by NASA under the LWS (grant \# 80NSSC19K0067), O2R (grant \# 80NSSC20K0285), and SBIR/STTR programs, by the NSF PREEVENTS program (ICER1854790), and AFOSR (contract \# FA9550-15-C-0001).  Computational resources were provided by NASA's NAS (Pleiades) and NSF's XSEDE (TACC \& SDSC).
\added{Simulation output files containing the data for all streams shown in Figures \ref{fig:streams-3D-int_flux} and \ref{fig:streams-3D-physics}, as well as interactive figure panels, are archived at \url{https://doi.org/10.5281/zenodo.4321562}.}
\added{The authors thank the reviewer for thoughtful comments and suggestions.}
%

\vspace{5mm}








\bibliography{bastille}
\bibliographystyle{aasjournal}


\listofchanges

\end{document}